\documentclass[twoside,fleqn]{article}
\usepackage{epsf}
\usepackage{espcrc2}
\usepackage{amssymb}
\usepackage{amsfonts}

\newcommand{\beq}{\begin{equation}}
\newcommand{\eeq}{\end{equation}}
\newcommand{\beqn}{\begin{eqnarray}}
\newcommand{\eeqn}{\end{eqnarray}}
\newcommand{\beqns}{\begin{eqnarray*}}
\newcommand{\eeqns}{\end{eqnarray*}}

\newcommand{\al }{\alpha}

\newcommand{\be}{\begin{equation}}
\newcommand{\ee}{\end{equation}}
\newcommand{\ba}{\begin{eqnarray}}
\newcommand{\ea}{\end{eqnarray}}
\newcommand{\bea}{\begin{eqnarray*}}
\newcommand{\bet}{\begin{center} \begin{tabular}}
\newcommand{\eea}{\end{eqnarray*}}
\newcommand{\ent}{\end{tabular} \end{center}}
\newcommand{\bb}{\begin{thebibliography}}
\newcommand{\eb}{\end{thebibliography}}

\newcommand{\bit}{\begin{itemize}}
\newcommand{\eit}{\end{itemize}}

\newcommand{\MSb}{$\overline{\mathrm{MS}}$ }

\def\lsim{\mathrel{\rlap{\lower4pt\hbox{\hskip1pt$\sim$}}
    \raise1pt\hbox{$<$}}}                % less than or approx. symbol
\def\gsim{\mathrel{\rlap{\lower4pt\hbox{\hskip1pt$\sim$}}
    \raise1pt\hbox{$>$}}}                % greater than or approx. symbol

\def\3{\ss}

\title{ \vspace{-3.65cm}                                     % 
       \vspace{1.78cm}                                      % for 3 preprint #
       The Vacuum Polarization: Power Corrections beyond OPE ?
            \thanks{Talk given by W. K\"urzinger at Lattice 2000, Bangalore, India.}}                     % for preprint

\author{M.~G\"ockeler%
           \address{Institut f\"ur Theoretische Physik, Universit\"at
                    Regensburg, D-93040 Regensburg, Germany},
        R.~Horsley%
           \address{Deutsches Elektronen-Synchrotron DESY \& NIC,
                    D-15735 Zeuthen, Germany}
                       \hspace{-0.27cm} $^,$\hspace{-0.12cm}
           \address{Institut f\"ur Physik, Humboldt-Universit\"at zu Berlin,
                    D-10115 Berlin, Germany},
        W.~K\"urzinger$^{\rm b,d}$,
        V.~Linke%
        \address{Institut f\"ur Theoretische Physik,
                    Freie Universit\"at Berlin, D-14195 Berlin, Germany},
        D.~Pleiter$^{\rm b,d}$,
        P.~E.~L. Rakow$^{\rm a}$ and
        G.~Schierholz$^{\rm b,}$%
           \address{Deutsches Elektronen-Synchrotron DESY,
                    D-22603 Hamburg, Germany}
}

\begin{document}

\begin{abstract}
%We compute the Adler function from the product of electromagnetic 
%currents in quenched QCD. The result is compared with predictions
%of perturbation theory and the operator product expansion. In the 
%former case we obtain the running coupling constant ${\alpha}_s$.
We compute the vacuum polarization on the lattice using
non-perturbatively $O(a)$ improved Wilson fermions. The result is
compared with the operator product expansion (OPE). 
\end{abstract}

\maketitle

% reset footnote counter
\setcounter{footnote}{0}

%----------------------------------------------------------------------------

\section{INTRODUCTION}
\label{intro}

\begin{figure}[htb] 
\hbox{
\epsfysize=3.5cm
\epsfxsize=7.cm
\epsfbox{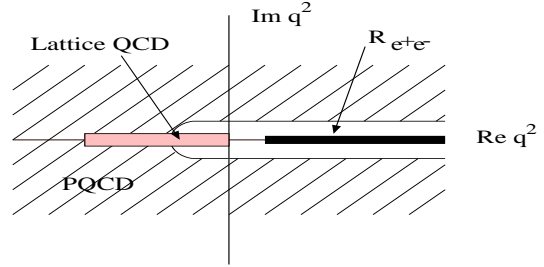}}
\caption{\label{KKK}\it{The analytic domain for the Adler function. 
Perturbative QCD is applicable
in the hatched area of the complex $q^2$-plane while lattice data
can provide information in the shaded region.
The picture is taken from \cite{Bjorken:2000ni}.}}
\end{figure}

The vacuum polarization is given by
\beqn
\label{vacuum_tensor1}
   \Pi_{\mu \nu}(q)\!\!\!\!\!\!\!\!\!\!  && = i \int d^4 x \; e^{iqx}\! 
\langle
0|T J_\mu (x) J_\nu(0) |0
\rangle
 \nonumber \\
    \hspace{0.0cm} && = (q_\mu q_\nu-q^2g_{\mu\nu})\,\Pi(q^2), 
\eeqn
with the vector current
$
J_{\mu}(x) =
\bar{\psi} (x)\gamma_{\mu}
 \psi (x) $. 
In the following we work at the 
Euclidean momenta
$Q^2 = -q^2$.
Because the polarization function $\Pi(-Q^2)$
has logarithmic divergences, it is customary to study the Adler
function \cite{Adler:1974gd}
\beq
\label{vacuum_tensor12}
D(Q^2) = - 12 \pi^2 Q^2\frac{d\Pi(-Q^2)}{dQ^2}\, .
\eeq
The Adler function provides 
a way of comparing theoretical predictions 
from QCD with available time-like experimental data for 
the $e^{+} e^{-}$ total cross section via
\begin{equation}
D(Q^2)=Q^2\int_{4 m_{\pi}^2}^{\infty}\frac{  R(s)}{(s+Q^2)^2}ds\; ,
\label{Adler}
\end{equation}
where 
\begin{equation}
R(s)=\frac{\sigma(e^+e^-\rightarrow
  {\rm hadrons})}{\sigma(e^+e^-\rightarrow \mu^+\mu^-)}
\end{equation}
is the ratio of the hadronic to the leptonic cross section 
at center of mass energy squared $s$.
%The leptonic cross section is
%$\sigma(e^+e^-\rightarrow \mu^+\mu^-)=\frac{4\pi\alpha^2}{3s}$
%where $\alpha=e^2/4\pi$ is
%the QED fine structure constant.
Lattice calculations will give valuable information
for the Adler function
in the region shown in 
Figure~\ref{KKK}.

The standard operator product expansion (OPE) of the
Adler function is
of the form
%given by
%corrections
\ba
\label{opes}
D(Q^2)=D^{\rm{pert}}(Q^2)+D^{\mathrm{NP}}(Q^2)\, .
\ea
%
%Because there is no non-vanishing local dimension two operator
%in the chiral limit, the lowest dimension operator in the
%non-perturbative part, the gluon condensate $\langle
%\frac{\alpha_s}{\pi} G G \rangle$  is of dimension 
%four. In the massive case we also have contributions from the
%dimension-four chiral condensate $\langle m_q \bar{q}q \rangle$.
The perturbative part is available to three loops in the form of a large-$Q^2$ expansion \cite{Eidelman:1998vc,Chetyrkin:1996cf,Chetyrkin:1997qi} and 
reads
%Therefore the  OPE of the Adler function in the large $|Q^2|$-limit reads
%
\begin{eqnarray}
\lefteqn{
\label{wilson1}
D^{\rm{pert}}(Q^2) =
3\sum_q Q_q^2 } \nonumber  \\
\lefteqn{
\left.
\Bigg\{
1\!-\!6\frac{m_q^2}{Q^2\!}-\!
12\frac{m_q^4}{Q^4} \ln
\frac{m_q^2}{Q^2}
%\right.}
% \nonumber  \\
%\lefteqn{
%\left.
\!+\!24\frac{m_q^6}{Q^6} 
\left(\!\ln
\frac{m_q^2}{Q^2}\!+\!1\!\right)
%+
%O \!
%\left(
%\frac{m_q^8}{Q^8}
%\right)
%\ln \frac{m_q^2}{Q^2}
\right.
}
\nonumber \\
%&&
%\left.
%24\frac{m_f^6}{Q^6} 
%\left(\ln\frac{m_f^2}{Q^2}+1\right) 
%\nonumber \\
\lefteqn{
\left.
+\frac{\al_s(\mu^2)}{\pi} 
\left[
1-12
 \frac{m_q^2}{Q^2} 
(
1
+
\ln\frac{m_q^2}{Q^2})
\right.
\right.
%\right)
}
\nonumber \\
\lefteqn{
\left.
\left.
+16 \frac{m_q^4}{Q^4}
\left(
\frac{17}{24}+2\zeta(3)
\!-\!
\frac{1}{4}
\ln\frac{m_q^2}{Q^2}
\!-\!
\frac{3}{2}
\ln^2\frac{m_q^2}{Q^2}
\right)
\right.
\right.
%\right)
}
\nonumber \\
\lefteqn{
\left.
\left.
-64 \frac{m_q^6}{Q^6}
\left(
\frac{139}{108}
-
\frac{19}{9}
\ln\frac{m_q^2}{Q^2}
-\frac{29}{24}
\ln^2\frac{m_q^2}{Q^2}
\right)
\right.
\right]
}
\nonumber \\
\lefteqn{
+
O \!
\left(
\frac{m_q^7}{Q^7}
\right)
%\ln \frac{m_q^2}{Q^2}
%}
%\nonumber \\
%\lefteqn{
%\left.
+
O \!
\left(
\al_s^2
\right)
% \right.
% \nonumber \\
% &&
% \left.
% \left(
% \frac{\al_s(\mu^2)}{\pi} 
% \right)^2 \ldots
\Bigg\}
 \, .}
%\sum\limits_{q=u,d,s} Q_q^2 N_{cq}\,
\end{eqnarray}
To leading order in $1/Q^2$ 
%The lowest dimensional operators of 
the non-perturbative part is given by 
%
%\beqn
\begin{eqnarray}
\label{wilson2}
\lefteqn{
%&&D(Q^2) =
D^{\mathrm{NP}}(Q^2)=
3\sum_q Q_q^2
8\pi^2
%}
%\nonumber \\
%&&
%\lefteqn{
}
 \\
\lefteqn{
\left\{
w_1
\frac{
\langle
\frac{\alpha_s}{\pi} G G
\rangle
}{Q^4} 
+w_2
\frac{
\langle
m_q \bar{q}q
\rangle
}{Q^4}
\right.
}
\nonumber \\
\lefteqn{
\left.
+w_3
\sum\limits_{q'} 
\frac{
\langle
m_{q'} \bar{{q'}}{q'}
\rangle
}{Q^4}\,
%+  \mbox{higher condensates}
%+
% O \!
%\left(
%\al_s^3
%\right)
%\ldots
+ \mbox{higher condensates}
%+\mbox{higher dim. Operators}
\,\right\}.
\nonumber 
} 
\end{eqnarray}
%\eeq
%
%
The last sum in eq.~(\ref{wilson2}) is over $q'$ dynamical quark flavors
which will not concern us here since we are working in the quenched approximation.
The Wilson coefficients are given by  \cite{Eidelman:1998vc}
\begin{eqnarray}
\label{w_1}
\lefteqn{
w_1=
\frac{1}{12} 
\left(1-\frac{11}{18}\frac{\alpha_s(\mu^2)}{\pi} \right) +O \!
\left(
\al_s^2
\right)
}
\\
%&&
\lefteqn{
w_2=
%2
%\left[
2+
\frac{2}{3}
\frac{\alpha_s(\mu^2)}{\pi}
%\right.
}
\nonumber 
\\
\lefteqn{
\label{w_2}
%\left.
\left(\frac{47}{4}-\frac{3}{2}
\ln{
\frac{Q^2}{\mu^2}}\right)
%\right]
\left(\frac{\alpha_s(\mu^2)}{\pi} \right)^2
+
O \!
\left(
\al_s^3
\right)
%\ln\frac{\mu^2}{Q^2}
}
 \\
%&&
\lefteqn{
\label{w_3}
w_3=
\frac{4}{27}\frac{\alpha_s(\mu^2)}{\pi}
} 
 \\
%&&
+\!
\lefteqn{
\left(\frac{4}{3}\zeta_3
\!
-
\!
\frac{88}{243}
\!
-
\!
\frac{1}{3}
\ln{\frac{Q^2}{\mu^2}}
\right)\!
%\right.
%\right.
%}
%\nonumber \\
%&&
%\lefteqn{
%\left. 
%\left. 
%\times
\left(\frac{\alpha_s(\mu^2)}{\pi} \right)^2
\!\!\!
+
O \!
\left(
\al_s^3
\right) .
} 
\nonumber
\end{eqnarray}
%&&
%\eeqn
%
%
%
Computation of the vacuum polarization allows us in principle to
determine the strong coupling constant $\alpha_s$ as well as the gluon, chiral and higher condensates.

It has been claimed by several groups \cite{Chetyrkin:1999yr,Zakharov:1998xs,Akhoury:1998hi,Vainshtein:1994ff} 
that the Adler function (as well as the polarization tensor) receives a further contribution of
the form
\beqn
\label{renorm}
\delta 
D(Q^2)   
\sim
{\Lambda^2}/{Q^2} \, .
\eeqn
Such a term is not present in the OPE,
because there exists no gauge invariant
dimension-two operator.
In this talk we shall compute the vacuum polarization and compare it
with the predictions of the OPE.
\section{LATTICE EVALUATION}
To construct the polarization tensor on the lattice
we demand the lattice Ward identity to be fulfilled.
In lattice momentum space this leads to
\beqn
\label{ward}
\hat{q}_\mu \, \Pi_{\mu \nu}=0 
\eeqn 
where the lattice momenta are defined as
$\hat{q}_\mu=(2/a)\sin(q_\mu a/2)$. 
(All momenta are Euclidean.)
The polarization tensor splits into two parts:
\beqn
\label{ward2}
\Pi_{\mu \nu} = \Pi_{\mu \nu}^{(a)}+\Pi_{\mu \nu}^{(b)}.
\eeqn 
The first part % of the tensor eq.~(\ref{ward2}) 
is given by 
\begin{figure}[htb] 
\hbox{
\epsfysize=7.3cm
\epsfxsize=6.5cm
\epsfbox{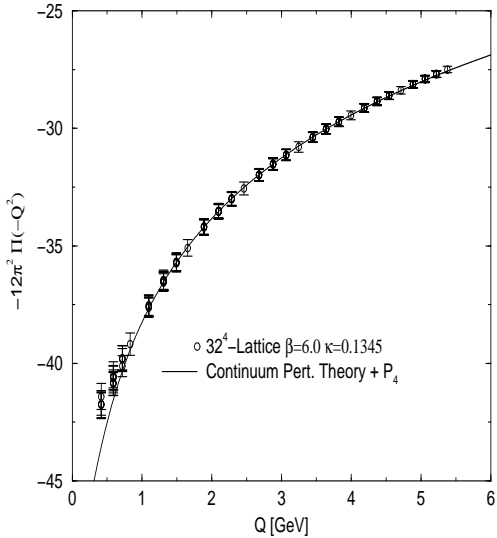}}
\caption{\label{lati}\it{The lattice data for $\beta=6.0$, $\kappa=0.1345$ on a $32^4$ lattice.
The line is the perturbative contribution. 
The scale is set using the $r_0$-parameter 
\cite{Guagnelli:1998ud}.}}
\end{figure}
\begin{figure}[htb] 
\hbox{
\epsfysize=7.3cm
\epsfxsize=6.5cm
%\epsfbox{xmgr_eins_durch_es_FINAL_SCH_KU_FETT_LL.eps}}
\epsfbox{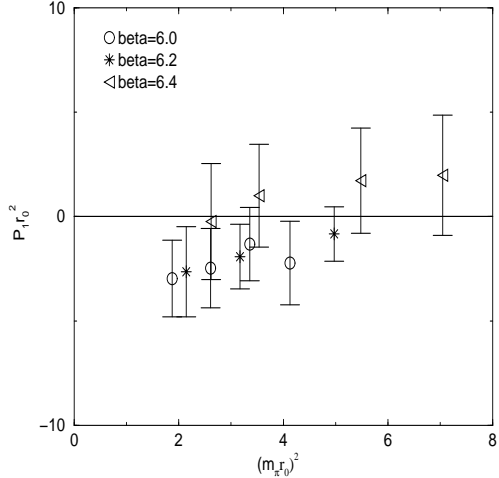}}
\caption{\label{latif}\it{The coefficient $P_1$
of the power correction $1/Q^2$.
The fit interval is chosen from approx. $1$~$\rm{GeV}$ to $5$~$\rm{GeV}$ for all lattices.}}
\end{figure}
\begin{eqnarray}
\label{diaa}
\lefteqn{\Pi_{\mu \nu}^{(a)}(q) =
\sum_{x}e^{iqx+iq\hat{\mu} a /2-iq\hat{\nu} a /2 }
}
\nonumber \\
&&\times
 \langle \, 0 \, |
J_{\mu}^{(1)}(x)  J_{\nu}^{  (1) \, +}(0) 
|\, 0\, \rangle \, ,
\end{eqnarray}
while the second part corresponds to
the tadpole diagram
\beqn
\label{diab}
\Pi_{\mu \nu}^{(b)}(q) =
\langle \, 0 \, |
J_{\mu}^{(2)}(0)  
|\, 0\, \rangle 
\delta_{\mu \nu}
\, .
\eeqn
For the current in eq.~(\ref{diaa}) we take 
the conserved vector current (CVC) 
\begin{eqnarray}
\label{CCV2}
\lefteqn{J_{\mu}^{(1)}(x)=
\frac{1}{2}(
\bar{\psi} (x+a \hat{\mu}) 
\left( 1+\gamma_{\mu}\right)
U_{\mu}^{+}(x)
\psi (x)} \nonumber \\ 
&&-\bar{\psi} (x) 
\left( 1-\gamma_{\mu}\right)
U_{\mu}(x)
\psi (x+a \hat{\mu}) )\, ,
\end{eqnarray}
and the current in 
eq.~(\ref{diab}) is given by
\begin{eqnarray}
\label{CCV3}
\lefteqn{
J_{\mu}^{(2)}(x)= 
\frac{1}{2}  (
\bar{\psi} (x+a \hat{\mu}) 
\left( 1+\gamma_{\mu}\right)
U_{\mu}^{+}(x)
 \psi (x) )} \nonumber \\ 
&&
+
\bar{\psi} (x) 
%&\! \! {{+}} & \! \! \bar{\psi} (x) 
\left( 1-\gamma_{\mu}\right)
U_{\mu}(x)
\psi (x+a \hat{\mu}) )\, .
\end{eqnarray}
We shall take the following ansatz for the polarization tensor:
\beqn
\Pi_{\mu \nu}=  
(\hat{q}_\mu \hat{q}_\nu
-\hat{q}^2 \delta_{\mu\nu}
)\,\Pi(-\hat{q}^2 )\, .
\eeqn

The calculations are done for non-pertur\-batively 
$O(a)$ improved Wilson fermions. Besides improving the fermionic action we have to improve the current 
as well.  The  on-shell improved conserved vector current is
\begin{eqnarray}
\lefteqn{
\label{CCV}
J_{\mu}^{(1) \, \rm{imp}}(x)
=
J_{\mu}^{(1)}(x)}
\nonumber  \\
& & 
+
\frac{c_{\rm cvc}}{2}
a
i
%\sum_{\lambda}
\partial_{\lambda}
%\delta^{\mu}_{\lambda}
\left\{
\bar{\psi} (x)
\sigma_{\mu \lambda }
\psi (x)
\right\}\, .
\end{eqnarray}
The value of the improvement coefficient $c_{\rm cvc}$
is not known beyond tree level. 
%at higher orders.
We take the tree level value 
%In the following we shall thus work with the tree level result 
$c_{\rm cvc}=1$.
%and not is known for the interacting case.
%The derivative is taken as 
The derivative in eq.~(\ref{CCV})
must be defined such that the Ward identity, eq.~(\ref{ward}),
is fulfilled.
We take
\beqn
\label{abl}
\partial_{\lambda}
f(x)=
\frac{1}{4a}\{
f(x+a\hat{\lambda})
-f(x-a\hat{\lambda})
+\nonumber\\
f(x+a\hat{\mu}+a\hat{\lambda})
-f(x+a\hat{\mu}-a\hat{\lambda})\}\, .
\eeqn 
%
%We calculate the matrix element in eq.~(\ref{diaa}) using
%this improved vector current.
%The derivative in eq.~(\ref{abl}) 
%gives in 
%momentum space 
% eq.~(\ref{diaa})
%a factor 
In the momentum space this gives a factor
%
%\beqn
%\label{fff}
$ -i \hat{q}_\lambda \cos(a {q}_\lambda /2 )  \cos(a {q}_\mu/2)\,$
%\eeqn 
%
which results in large $O(a^2)$
corrections, even in the free case.
In the following we omit the
$O(a^2)$
terms,
thus leaving us with the factor
%working with the factor
%
%\beqn
%\label{fff1}
$ -i \hat{q}_\lambda$.
% \,.
%\eeqn 
%
This keeps the Ward identity %eq.~(\ref{ward})
fulfilled.
\section{PRELIMINARY RESULTS} 
We have performed simulations at the 
$\beta$
and
$\kappa$
values shown in Table \ref{SERES}.
Because the Adler function involves a
derivative 
with respect to $Q^2$
we prefer to work with the polarization function.
%We work with a fit function of the form
The polarization function is thus written as
\begin{table}
\begin{center}
\begin{small}
\begin{tabular}{ |l|l|l|l| }
 \hline
 $\beta$ &   $\kappa$  &    $V$         \\\hline\hline
   6.0   &   0.1336    &    $16^4$      \\\hline
   6.0   &   0.1339    &    $16^4$      \\\hline
   6.0   &   0.1342    &    $16^4$      \\\hline
   6.0   &   0.1345    &    $16^4$      \\\hline\hline
   6.0   &   0.1345    &    $32^4$      \\\hline\hline
   6.2   &   0.1344    &    $24^4$      \\\hline
   6.2   &   0.1349    &    $24^4$      \\\hline
   6.2   &   0.1352    &    $24^4$      \\\hline\hline
   6.4   &   0.1343    &    $32^4$      \\\hline
   6.4   &   0.1346    &    $32^4$      \\\hline
   6.4   &   0.1350    &    $32^4$      \\\hline
   6.4   &   0.1352    &    $32^4$      \\\hline
\end{tabular}
\end{small}
\end{center}
\caption{\itshape{The values of $\beta$ and $\kappa$ used in the simulations.}\label{SERES}}
\end{table}
\beqn
\label{fit}
\lefteqn{
-12 \pi^2
\Pi(-Q^2)
=
c_0 + c_1\,\alpha_s(\mu^2)
+ c_2\,\alpha^2_s(\mu^2) }
% \ldots
\nonumber \\
&&+ 
\Bigg\{
{{\frac{P_1}{Q^2}}}
+ 
{{\frac{P_2}{Q^4}}} 
+{P_3}{a^2Q^2}+P_4
\Bigg\} \,.
\eeqn
We have included an additive constant ($P_4$) 
to account for
the logarithmically divergent contribution \cite{Chetyrkin:1996cf,Kawai:1981ja}.
This depends on the renormalization scheme.
For the perturbative 
contribution we take the three-loop result
\cite{Chetyrkin:1996cf} renormalized in the \MSb scheme.
The first coefficient in eq.~(\ref{fit}) 
reads for a single quark with  charge $Q_q=1$
\beqn
\lefteqn{
c_0=-\frac{9}{4}\left[
         \frac{20}{9} -\frac{4}{3}\ln\frac{Q^2}{\mu^2}
         -8\frac{m^2}{Q^2}
\right.
}
\nonumber\\
&&         
\left.
+\left(\frac{4m^2}{Q^2}\right)^2
          \left(\frac{1}{4}+\frac{1}{2}\ln\frac{Q^2}{m^2}\right)
         + \ldots
       \right] 
\eeqn
and is related to the leading part in eq. (\ref{wilson1})
via a derivative with respect to $\ln(Q^2)$.
For $\alpha_s(\mu^2)$ we take the four-loop result \cite{vanRitbergen:1997va,Vermaseren:1997fq} with
$\Lambda_{\overline{\mathrm{MS}}}=238(19)\rm{MeV}$ \cite{Capitani:1999mq}.
The quark mass renormalizations are taken from \cite{Gockeler:2000cy}.
Thus the perturbative part is completely known.
We also allow for a $1/Q^2$ contribution.
Residual $O(a^2Q^2)$ corrections are accounted for by $P_3$.
In Figure~\ref{lati} we show
the lattice data for $\beta=6.0$ and $\kappa=0.1345$.
This is compared with the perturbative part of 
eq.~(\ref{fit}) including a constant term $P_4$.
There we have also set $\mu=1/a$.
However the result was found not to depend significantly on the exact choice of $\mu$.
For values of $Q^2>2 \rm{GeV}^2$ we find very good agreement between the lattice data
and three-loop perturbation theory.
Alternatively
fitting eq.~(\ref{fit}),
our data in $1 \rm{GeV}\lesssim Q \lesssim 5 \rm{GeV}$
% to the data
gives the result shown in Figure~\ref{latif}.
It turns out that
%The parameter 
$P_1$ is consistent with zero.
(Note that $r_0^2\approx 6/{\rm GeV}^{2}$.)
At present we are not able to quote a reliable number for $P_2$.
%It is not clear yet, if 
%systematical errors will allow us to determine gluon and chiral condensates reliably. 
%
%\section{CONCLUSIONS AND OUTLOOK}
%\label{conclusions}
%The investigation is still in progress. 
%We hope to better understand systematical errors
%and improve the statistics.
%nresolved problems are the size of the improvement coefficient
%c_{\rm cvc}$ beyond tree level, as well as the
%O(a^2)$ ambiguities and the low statistic. 
%resently it seems that the continuum OPE is consistent with the
%attice data                         .
%
\section*{ACKNOWLEDGEMENTS}
\label{acknowledgement}
The numerical calculations were performed on the
Quadrics parallel computer of DESY Zeuthen.
We thank the operating stuff for support.
%APE $QH1$, $QH2$ and $QH4$ at DESY Zeuthen.
%

\end{document}